\documentclass[12pt]{article}
\usepackage{epsf}
\def\be{\begin{equation}}
\def\ee{\end{equation}}
\def\bea{\begin{eqnarray}}
\def\eea{\end{eqnarray}}

\def\rcn{ R^{c/n}}

\begin{document}
\begin{titlepage}
\begin{center}
{\Large \bf William I. Fine Theoretical Physics Institute \\
University of Minnesota \\}
\end{center}
\vspace{0.2in}
\begin{flushright}
FTPI-MINN-13/37 \\
UMN-TH-3308/13 \\
October 2013 \\
\end{flushright}
\vspace{0.3in}
\begin{center}
{\Large \bf Isospin violation in the yield of S-wave heavy meson pairs near threshold
\\}
\vspace{0.2in}
{\bf Xin Li$^a$  and M.B. Voloshin$^{a,b,c}$  \\ }
$^a$School of Physics and Astronomy, University of Minnesota, Minneapolis, MN 55455, USA \\
$^b$William I. Fine Theoretical Physics Institute, University of
Minnesota,\\ Minneapolis, MN 55455, USA \\
$^c$Institute of Theoretical and Experimental Physics, Moscow, 117218, Russia
\\[0.2in]

\end{center}

\vspace{0.2in}

\begin{abstract}
We consider production of overall neutral pairs consisting of heavy $D^{(*)}$ or $B^{(*)}$ meson and antimeson in the processes, where the pair can be created in an $S$-wave, such as $e^+e^- \to  \pi^0 \, D^* \bar D^{(*)}$ and $e^+e^- \to \gamma \,  D^{(*)} \bar  D^{(*)}$, and similar reactions with $B^{(*)}$ mesons. The ratio $R^{c/n}$ of the yield of pairs of charged mesons to that of neutral mesons near the threshold for the heavy pair is strongly affected by the isospin breaking due to the isotopic mass differences and due to the Coulomb interaction between the charged mesons. The actual behavior of the isospin breaking in $R^{c/n}$ is also sensitive to the strong interaction between the heavy mesons, so that experimental measurement of this ratio can be used as a probe of this strong interaction. We calculate $R^{c/n}$ at and very near the threshold in terms of the isotopic scattering lengths for the meson pairs. In particular we find that the yield of pairs of charged mesons does not go to zero at exactly the threshold, but rather starts with a finite step, whose height depends on the value of the scattering length in a particular channel. 
\end{abstract}
\end{titlepage}

Recent experimental data have revealed existence of a considerable number of peaks near the thresholds for pairs of mesons containing heavy $c$ or $b$ quarks. Such are the $Z_b(10610)$ and $Z_b(10650)$ resonances~\cite{bellez} at respectively the $B B^*$ and $B^* B^*$ thresholds in the bottomonium sector, and the $X(3872)$ and the recently observed peaks $Z_c(3900)$~\cite{besz39,bellezc}, $Z_c(3885)$~\cite{besz3885}, $Z_c(4020)$~\cite{besz4020} and $Z_c(4025)$~\cite{besz4025} near the $D D^*$ and $D^* D^*$ thresholds. Some, if not all, of these peaks are related to the strong-interaction dynamics in the $S$-wave of a heavy meson-antimeson pair. It is thus highly likely that this dynamics will present a feature rich field for further experimental and theoretical studies. The $S$ wave states of  meson-antimeson pairs are not directly produced in $e^+e^-$ annihilation, but are still observable after emission of a pion or a photon with the added benefit that  spectra of invariant masses for the heavy meson pairs can be observed at a fixed energy of the $e^+e^-$ beams. These spectra can be studied practically down to the threshold, where the $S$-wave production dominates the yield. Clearly, the emission of pion produces the meson pair in a different isotopic and charge parity state than emission of a photon, so that a number of possible channels can be studied, even if the orbital motion is restricted to only one partial wave.

In this paper we address the charged-to-neutral ratio $\rcn$ of the yield of heavy meson pairs in the processes of the type $e^+e^- \to \pi^0 \, D^* \bar D^{(*)}$ and $e^+e^- \to \gamma \,  D^{(*)} \bar  D^{(*)}$ and similar processes with the $B^{(*)}$ mesons very near and exactly at the threshold for the corresponding pair of charged mesons. The deviation of this ratio from the value, determined by the isotopic properties of the source of the pairs, due to the isospin-violating mass differences within the isotopic doublets of heavy mesons and due to the Coulomb interaction between the charged mesons is most significant near the threshold~\footnote{Clearly the isotopic mass difference is not essential for the $B$ and $B^*$ mesons where it is known to be very small, but the Coulomb interaction gives rise to a significant effect in $\rcn$~\cite{mv12}.}. 
Furthermore, it has been argued previously~\cite{mv03,dlorv,mv12} that the specific expressions for the isospin-violating effects in the charged-to-neutral yield ratio are sensitive to the strong-interaction scattering phases and can thus serve as a probe of the force between the heavy mesons.  The formulas for the dependence of $\rcn$ on the scattering phases were found for a $P$-wave~\cite{dlorv} and for an $S$-wave~\cite{mv12} production in the first order in the isospin-violating mass and electric charge differences. Given that the effect of these differences grows as the energy decreases towards the threshold, the  available expressions are not applicable very near or at the threshold where this effect is the largest and, hopefully, is more readily measurable. For this reason we derive here the expressions for $\rcn$ that are valid to all orders in the isotopic mass difference and in the Coulomb interaction between the charged mesons. Our treatment is complementary to the previous studies in that it is applicable at low energy above the threshold in the center of mass of the heavy meson pair, where the scattering between the mesons can be described in the $S$-wave within the small interaction radius approximation~\cite{ll} in terms of the scattering lengths $a_0$ and $a_1$ in the channels with definite isospin. We find in particular that due to the Coulomb attraction the yield of pairs of charged mesons does not go to zero at exactly the threshold, but rather starts with a finite step, at which the ratio $\rcn$ for charmed mesons is generally comparable to one with the specific value being determined by the isotopic mass difference and the appropriate strong scattering length. 

In what follows we start with describing the approach to the problem and first derive the expression for $\rcn$ in the case where the meson pair is produced by an isotopically pure source, such as e.g. $I=1$ source in  $e^+e^- \to \pi^0 \, D^* \bar D^{(*)}$. We then consider the case of an isotopically mixed source such as in the process $e^+e^- \to \gamma \,  D^{(*)} \bar  D^{(*)}$ where the photon can be emitted by the current of light quarks. 

In the small interaction radius approximation (see e.g. in the textbook~\cite{ll}) it is assumed that the strong interaction is limited to distances $r$ between the mesons shorter than an effective radius $r_0$.  At $r < r_0$ the potential for the strong interaction is assumed to be much larger than the isospin violating terms and also much larger than the variation of the center of mass energy $E$ from the threshold in the range where the discussed approximation is considered. The former assumption implies that the interaction at $r < r_0$ depends only on the isospin, and that in this `inner' region the system with a fixed orbital momentum is described by the radial wave functions with definite isotopic spin: $\phi_0(r)$ for $I=0$ and $\phi_1(r)$ for $I=0$. In the absence of any sources these functions should be regular at $r=0$. We use throughout this paper the notation $\phi(r) = r \, R(r)$ with $R(r)$ being the radial part of the solution of the three-dimensional Schr\"odinger equation in a given partial wave, so that in the $S$-wave the regularity implies $\phi_{0,1}(0)=0$. These solutions to the `inner' problem are to be matched to the `outer' wave functions at $r=r_0$, so that the matching conditions are determined by their logarithmic derivatives at $r_0$:
\be
   \phi_0'(r)/\phi_0(r)|_{r=r_0} = -\kappa_0, ~~~~~ \phi_1'(r)/\phi_1(r)|_{r=r_0} = -\kappa_1.
\label{kappas}
\ee
In the energy range that is much smaller than the potential of the strong interaction in either of the isotopic channels the dependence of the wave functions $\phi_0$ and $\phi_1$ on the specific value of the energy can be neglected, so that the constants $\kappa_{0,1}$ can be considered as independent of the energy.  
At $r > r_0$ the strong force can be entirely neglected, and the wave functions for the meson pairs are described either by a free particle Schr\"odinger equation for neutral particles, or by the motion in the Coulomb potential for the charged mesons. The solutions to these `outer' equations are to be used for matching at $r=r_0$ the logarithmic derivatives in Eq.(\ref{kappas}). In the limit where the mass differences and the Coulomb interaction are neglected, the isotopic symmetry is exact for the outer problem as well, and the  solutions for the outer problem are simply the plane waves $\exp ( \pm i p r)$ with $p$ being the momentum of each of the mesons in the center of mass frame. Furthermore, at low energy above the threshold the momentum is also small, so that $p r_0 \ll 1$ and (for the `outer' functions) the matching conditions can be considered as shifted from $r=r_0$ to $r=0$. In this limit one thus readily finds the well known expressions for the scattering amplitudes in the channels with definite isospin~\cite{ll}
\be
f_0=-{1 \over \kappa_0 + i \,p},~~~~~~~~~~~~ f_1=-{1 \over \kappa_1 + i \,p}~,
\label{fs}
\ee
so that the scattering length [defined as $-f(p \to 0)$] in each channel is the inverse of the corresponding $\kappa$:
\be
a_{0,1} = {1 \over \kappa_{0,1}}~.
\label{as}
\ee

The expressions (\ref{fs}) correspond to the scattering phases $\delta_0$ and $\delta_1$ in the  isotopic channels given as
\be
\cot \delta_{0,1} = - {\kappa_{0,1} \over p}~.
\label{ds}
\ee 
This relation shows the deficiency of using the scattering phases for discussion of isospin violating effects beyond the first order. Indeed, once e.g. the isotopic mass difference between the mesons is taken into account, the momentum $p$  for the pair of neutral mesons, $p_n$, is different from that for the pair of charged ones, $p_c$. As a result the notion of the scattering phases for definite isospin becomes ambiguous, and this ambiguity is determined by the ratio of the mass difference to the excitation energy above the threshold. On the contrary, the isotopic parameters $\kappa_{0,1}$ (or, equivalently, the scattering lengths) are stable as long as the isospin violating terms are small in comparison with the energy of the strong interaction in the `inner' region and can be neglected. (A further discussion of the scattering amplitudes for the $D \bar D^*$ meson pairs in this approximation with the isotopic mass difference taken into account can be found in Ref.~\cite{mv07}.)

Proceeding to our derivation of the formulas for $R^{c/n}$ we start with neglecting the Coulomb effect and considering only the isospin violation by the isotopic mass difference in a process, where the heavy meson pair is produced by an isotopically pure source. For definiteness we consider a source producing the heavy meson pairs in the $I=1$ state, as is the case, we believe to a good accuracy, for the processes e.g. $e^+e^- \to \pi^0 D \bar D^*$ and $e^+e^- \to \pi^0 D^* \bar D^*$. Unlike in the scattering problem, the relevant for the production process wave functions contain only outgoing waves. Denoting $\phi_c(r)$ ($\phi_n(r)$) the `outer' wave function for the pair of charged (neutral) mesons, one can formulate the problem as that of finding the solution that up to an overall normalization factor (common for $\phi_n$ and $\phi_c$) at $r > r_0$ reads as
\be
\phi_c(r) = b_1 \, \exp(i p_c r), ~~~~~~~~~~ \phi_n(r) = \exp(i p_n r)~,
\label{defb}
\ee
where the notation $b_1$ implies that this coefficient arises in a situation where the source is a pure isovector.
The ratio of the outgoing fluxes in the waves in Eq.(\ref{defb}) determines the ratio $\rcn$:
\be
\rcn = {p_c \over p_n} \, |b_1|^2~,
\label{rb}
\ee
so that the problem reduces to finding the coefficient $b_1$.

As different from the scattering problem, in the production process there is an isovector source coupled to the $I=1$ function $\phi_1$. We consider here the case where the source is located within the region of the strong interaction, i.e. at $r < r_0$. This appears to be a reasonable assumption for the practical experimental conditions. Indeed, the emission of the pion is a strong interaction process with the corresponding distance scale, and the only `smearing' of the source could arise for soft pions from the pion Compton wave length.  In the actual measurements at the $e^+e^-$ energy $\sqrt{s} \approx 4.26\,$GeV and above the pion momentum for the production of the charmed meson pairs at the threshold is at least as large as 0.2\,GeV, so that the corresponding characteristic distance is also comparable to the scale of the strong interaction. Thus the isovector `inner' wave function $\phi_1$ is not a solution of the `inner' problem without a source, and the second of the boundary conditions in Eq.(\ref{kappas}) should not be used. However the isoscalar channel is not affected at $r < r_0$ by the source, and the isoscalar function $\phi_0(r)$ is not changed and satisfies the first of the relations in Eq.(\ref{kappas}). Thus the isoscalar combination of the wave functions (\ref{defb}), $\phi_0 = \phi_c + \phi_n$ should still satisfy this boundary condition at $r = r_0$. Shifting, as before, the matching point to $r=0$, one readily finds the expression for the coefficient $b_1$:
\be
b_1 = - {\kappa_0 + i p_n \over \kappa_0+ i p_c}~,
\label{b1}
\ee 
so that 
\be
\rcn=  {p_c \over p_n} \, {\kappa_0^2 +  p_n^2 \over \kappa_0^2+  p_c^2}~.
\label{r1}
\ee

Including the effect of the Coulomb attraction between the charged mesons amounts to replacing the outgoing plane wave $\exp(i p_c r)$ in the `$c$' channel with the exact solution in the Coulomb potential $V_C=-\alpha/r$ which asymptotically at $r \to \infty$ is an (appropriately normalized) outgoing wave. This solution is well known (and can be found e.g. in the textbook~\cite{ll}) and is given by 
\be
g(r)= -2 i \, p_c r \, \exp(i p_c r) \, \left [ {  1- \exp (-2 \pi \lambda) \over 2 \pi  \lambda} \right ]^{1/2} \,  \Gamma (1- i \, \lambda)  \, U(1-i \, \lambda, 2, - 2 i \, p_c r)~,
\label{gc}
\ee
where $U(a,b,z)$ is the standard confluent hypergeometric function of the second kind, and $\lambda$ stands for the Coulomb parameter,
$\lambda = m \, \alpha/p_c$,
with $m$ being the reduced mass for the pair of charged mesons, e.g. $m \approx 1.005\,$GeV for the $D^{*+}  D^{*-}$ pair and $m \approx 0.97\,$GeV for  $D^+  D^{*-}$. The solution in Eq.(\ref{gc}) at large $r$ describes an outgoing wave with the flux normalized to $p_c$, while its expansion at small $r$ reads as
\be
g(r) = i \, \xi \, p_c r + {1 \over \xi} \, \left \{ 1 - 2 p_c r \lambda \, \left [ \log (2 p_c r) + 2 \gamma_E - 1 + {\rm Re}\, \psi(i \, \lambda) \right ] \right \} + O( r^2 \log r)~,
\label{gcs}
\ee
where $\gamma_E \approx 0.5772$ is the Euler constant, $\psi(z) = \Gamma'(z)/\Gamma(z)$ is the logarithmic derivative of the Gamma function, and $\xi$ is introduced for notational simplicity as
\be
\xi= \left [ {2 \pi  \lambda \over 1- \exp (-2 \pi \lambda)} \right ] ^{1/2}~,
\ee
so that $\xi^2$ is the well known Sommerfeld factor for the Coulomb attraction.

As it can be seen from Eq.(\ref{gcs}), the derivative of the real part of the wave function $g(r)$ has a logarithmic singularity at $r \to 0$. Therefore in the matching conditions the distance $r_0$ in this logarithmic term should be kept finite, while in the rest of the terms it can still be replaced by zero. The coefficient $b_1$ is then readily found from the matching conditions as
\be
b_1 =-  \xi \, { \kappa_0 + i p_n \over \kappa_0 - 2 m \alpha \, \left [ \log (2 p_c r_0) + 2 \gamma_E  + {\rm Re}\, \psi(i \, \lambda) \right ] + i p_c \xi^2 }~.
\label{b1f}
\ee
The final expression for the ratio $\rcn$ with a purely isovector source is thus given by
\bea
&&\!\! \! \! \! \! \rcn =  {p_c \over p_n} \, {2 \pi m \alpha / p_c  \over 1- \exp(-2 \pi m \alpha /p_c) } \, \times \label{r1f} \\ \nonumber 
&&\!\! \! \! \! \!  { \kappa_0^2 + p_n^2 \over \left \{ \kappa_0 - 2 m \alpha \, \left [ \log (2 p_c r_0) + 2 \gamma_E  + {\rm Re}\, \psi(i \, m \alpha /p_c ) \right ] \right \}^2 + p_c^2 \, (2 \pi m \alpha/p_c)^2/ [ 1- \exp(-2 \pi m \alpha/p_c) ]^2 }~.
\eea

The discussed treatment can be readily adapted for the case of an isoscalar source of the heavy meson pairs. Indeed, in this case
it is the isovector wave function $\phi_1$ which satisfies at $r < r_0$ the Schr\"odinger equation without source, so that the isovector combination $\phi_c - \phi_n$ of the wave functions in the two channels, describing at $r > r_0$ the outgoing waves,
\be
\phi_c(r) = b_0 \, g(r)~, ~~~~~~~~~ \phi_n(r)=\exp(i p_n r)~,
\label{defb0}
\ee
has to satisfy at $r=r_0$ the second boundary condition in Eq.(\ref{kappas}). Using the explicit expression (\ref{gcs}) for the Coulomb  modified wave function, one then readily finds
\be
b_0 =  \xi \, { \kappa_1 + i p_n \over \kappa_1 - 2 m \alpha \, \left [ \log (2 p_c r_0) + 2 \gamma_E  + {\rm Re}\, \psi(i \, \lambda) \right ] + i p_c \xi^2 }
\label{b0f}
\ee
and, accordingly, the ratio $\rcn$ is given by the same expression as in Eq.(\ref{r1f}) with $\kappa_0$ replaced by $\kappa_1$.

The expressions (\ref{b1f}) and (\ref{b0f}) are valid to all orders in the isotopic mass difference, i.e. the difference between $p_c$ and $p_n$ at a given energy, and in the Coulomb interaction between the charged mesons. It can be mentioned that
in the limit of a perfect isotopic symmetry ($p_c=p_n$ and $\alpha \to 0$) one finds $b_1=-1$, $b_0=1$ and $\rcn=1$ in both cases. If the isotopic mass difference is nonzero, but in the limit, where there is no strong interaction (formally corresponding to a zero scattering length, i.e. $\kappa \to \infty$) the usual phase space ratio $\rcn = p_c/p_n$ is recovered, provided that the Coulomb interaction is neglected. If the Coulomb attraction is accounted for in this limit, one recovers the well known Sommerfeld factor for the Coulomb enhancement of the production rate. At finite $\kappa_0$  the first order term of expansion of the expression (\ref{r1f}) in the isotopic mass difference and in $\alpha$ matches the previously known formula~\cite{mv12}, if the relation (\ref{ds}) for the phase shifts $\delta_{0,1}$ in terms of $\kappa_{0,1}$ is also used. Furthermore, in the discussed here approach it is quite natural that the isospin-violating effects in the ratio $\rcn$ for production by an isotopically pure source, i.e. with either $I=0$ or $I=1$, are influenced by the strong interaction in the orthogonal isospin channel, i.e. $I=1$ or $I=0$, respectively. This property, found in the first-order treatment~\cite{dlorv,mv12}, persists in all orders in the discussed isospin breaking terms, and can be used for studying the strong interaction between the heavy mesons in the isotopic states that may not be readily accessible~\cite{mv12}.

The general case of an isopically mixed source of heavy meson pairs can be considered using our previous formulas for the isotopically pure sources. Indeed, let $A_0$ and $A_1$ denote the production amplitudes for the meson pairs in the corresponding isotopic states. These amplitudes define the overall normalization factors multiplying the expressions (\ref{defb}) and (\ref{defb0}) for the outgoing waves at $r > r_0$:
\be
\phi_c(r) = (A_0 \, b_0 - A_1 \, b_1) \, g(r)~,~~~~~~~~~~~\phi_n(r) = (A_0 - A_1) \, \exp(i p_n r)~,
\label{defbm}
\ee
where the sign of $A_1$ is chosen for consistency with the conventions of Ref.~\cite{dlorv}. The full production amplitude is a linear superposition of two isotopic amplitudes. Thus the individual terms can be found by setting either $A_0$ or $A_1$ to zero, which reduces the problem to the case of isotopically pure source. Thus the coefficients $b_0$ and $b_1$ are exactly the ones given by the expressions (\ref{b1f}) and (\ref{b0f}), so that the ratio of the amplitudes of the waves in the `$c$' and `$n$' channels is given by
\be
b_m = {A_0 \, b_0 - A_1 \, b_1 \over A_0 - A_1}~,
\label{bmf}
\ee
and the ratio $\rcn$ of the fluxes in this situation is found as
\be
\rcn={p_c \over p_n} \, \left | {A_0 \, b_0 - A_1 \, b_1 \over A_0 - A_1} \right |^2~.
\label{rmf}
\ee

It is instructive to illustrate the significance of the discussed effects in $\rcn$ with numerical estimates. We present here such estimates for the case of the charmed meson pairs produced in the processes $e^+e^- \to \pi^0 + D^* \bar D^*$ and  $e^+e^- \to \pi^0 + (D \bar D^* + c.c.)$. In these reactions  the heavy meson pair is produced in the $I=1$ state, so that the charged-to-neutral yield ratio is determined by the strong interaction parameters $\kappa_0$ and $r_0$ according to the relation  (\ref{r1f}). In fact the dependence in Eq.(\ref{r1f}) on the effective radius $r_0$ is very weak as long as $r_0$ is much smaller than the `Bohr radius' for the system of $D^{(*)}$ mesons: $r_0 \ll 1/m \alpha \approx 27\,$fm, which is certainly the case. For this reason we fix $r_0$ at 1\,fm, and make estimates for a `representative' range of values for the parameter $\kappa_0$, which parameter is currently unknown, and which is generally different for $D^*\bar D^*$ and $D \bar D^*$ systems. The resulting  behavior of the ratio $\rcn$ near the threshold for $D^*\bar D^*$ is shown in Fig.~1. 

\begin{figure}[ht]
\begin{center}
 \leavevmode
    \epsfxsize=12cm
    \epsfbox{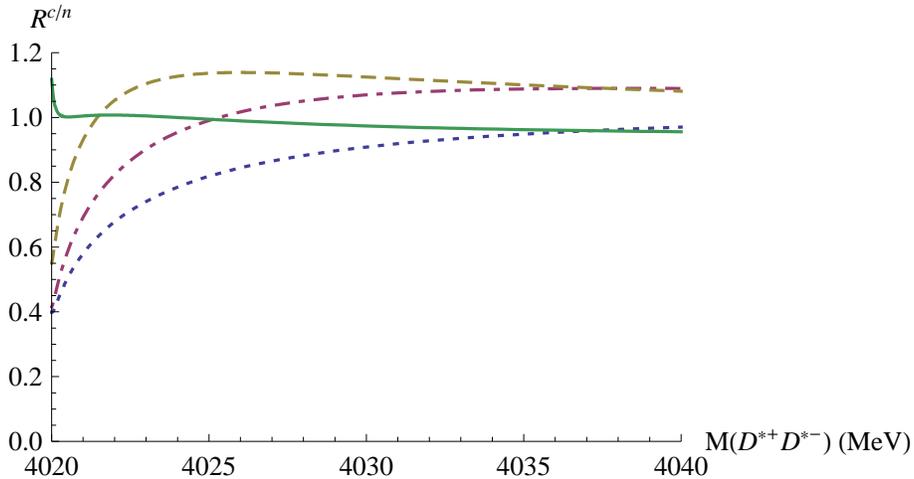}
    \caption{The charged-to-neutral yield ratio $\rcn$ for the $D^* \bar D^*$ pairs produced in $e^+e^- \to \pi^0  D^* \bar D^*$ near the threshold. The plots are calculated using Eq.(\ref{r1f}) with $r_0=1\,$fm and a set of different values of $\kappa_0$: -100\,MeV (solid), 100\,MeV (dashed), 200\,MeV (dot-dashed). The dotted curve is for the limit of no strong interaction between the mesons, formally corresponding to $\kappa_0 \to \infty$.}
\end{center}
\end{figure} 

\begin{figure}[ht]
\begin{center}
 \leavevmode
    \epsfxsize=12cm
    \epsfbox{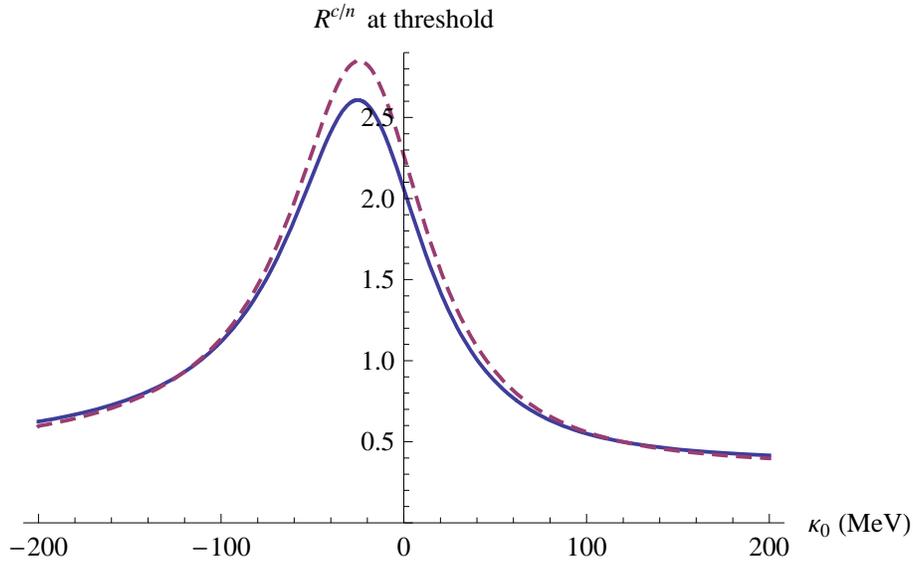}
    \caption{The dependence on $\kappa_0$ of the value of $\rcn$ at exactly the threshold for $D^{*+}D^{*-}$ (solid) and for $D^+D^{*-}$ ($D^-D^{*+}$) (dashed).}
\end{center}
\end{figure} 

The most characteristic feature, as readily seen from the plots, is that the ratio $\rcn$ does not go to zero at exactly the threshold in the invariant mass of the meson pair~\footnote{Clearly, this essentially due to the fact that the product $p_c \xi^2$ at $p_c \to 0$ approaches a finite value $2 \pi m \alpha \approx 45\,$MeV, which is not small in comparison with the momentum of the neutral pair $p_n$ at the same invariant mass (e.g. $p_n \approx 115\,$MeV for $D^{*0} \bar D^{*0}$ at the threshold of $D^{*+}
D^{*-}$).}, but rather starts with a finite step. The particular value of $\rcn$ at this point depends on $\kappa_0$  as illustrated in Fig.~2, and also weakly depends on the radius $r_0$, which is fixed at $r_0=1\,$fm in Fig.~2.
One can readily see that the behavior of the starting threshold values of $\rcn$ for the two types of the charmed meson pairs is very close. Thus any experimentally measured significant difference of these values would reveal a dissimilarity in the strong interaction between the isoscalar channels $D \bar D^*$ and $D^* \bar D^*$ with the quantum numbers $J^{PC}=1^{+-}$.

In summary. We have considered the interplay between the isospin violating mass differences, the Coulomb interaction and the strong scattering  in the threshold behavior of the charged-to-neutral yield ratio $\rcn$ for the $S$-wave production of overall neutral pairs of heavy $D^{(*)}$ or $B^{(*)}$ mesons. The expressions (\ref{b1f}), (\ref{b0f}) and (\ref{r1f}) for this behavior take into account all orders in the mass differences and in the Coulomb interaction and  are derived in the limit where the strong scattering of heavy mesons can be described by the scattering lengths in the isoscalar and isovector channels, which description is appropriate for the near-threshold behavior in the $S$ wave.  The  considered here production processes can be observed experimentally in the reactions such as   $e^+e^- \to  \pi^0 \, D^* \bar D^{(*)}$ and $e^+e^- \to \gamma \,  D^{(*)} \bar  D^{(*)}$, a study of which appears to be well within the capabilities of the current BESIII experiment. In particular we find that the onset of the yield of pairs of charged mesons starts with a finite step at the threshold. The height of the step is sensitive to a strong scattering length, and its measurement can be used as a probe of strong interaction between the heavy mesons. 

This work is supported, in part, by the DOE grant DE-FG02-94ER40823.

%\newpage

\end{document}